\documentclass[prd,twocolumn,amsmath,amssymb]{revtex4}
\usepackage{graphicx}

\begin{document}
\title{\boldmath Direct Measurements of the Branching Fractions for
the Semileptonic Decays $D^0 \to K^-\mu^+\nu_\mu$ and
$D^0 \to \pi^-\mu^+\nu_\mu$ }
\author{
M.~Ablikim$^{1}$,              J.~Z.~Bai$^{1}$,               Y.~Ban$^{12}$,
X.~Cai$^{1}$,                  H.~F.~Chen$^{16}$,
H.~S.~Chen$^{1}$,              H.~X.~Chen$^{1}$,              J.~C.~Chen$^{1}$,
Jin~Chen$^{1}$,                Y.~B.~Chen$^{1}$
% ...
\\
\vspace{0.2cm}
(BES Collaboration)\\
\vspace{0.2cm}
{\it
$^{1}$ Institute of High Energy Physics, Beijing 100049, People's Republic of China\\
% ...
$^{12}$ Peking University, Beijing 100871, People's Republic of China\\
% ...
$^{16}$ University of Science and Technology of China, Hefei 230026, People's Republic of China\\
% ...
}}

\begin{abstract}
Based on the data sample of 33 pb$^{-1}$ collected at and around 3.773 GeV
with the BES-II detector at the BEPC collider, the absolute branching
fractions for the semileptonic decays $D^0\to K^-\mu^+\nu_\mu$ and $D^0\to
\pi^-\mu^+\nu_\mu$ have been measured. In the system recoiling against
$7584\pm198\pm341$ singly tagged $\bar D^0$ mesons, $87.2\pm13.6$ events for
$D^0\to K^-\mu^+\nu_\mu$ and $9.3\pm7.4$ events for $D^0\to\pi^-\mu^+\nu_\mu$
are observed. These yield the absolute branching fractions to be
$BF(D^0\to K^-\mu ^+\nu_\mu)= (3.55\pm0.56\pm0.59)\%$ and
$BF(D^0\to\pi^-\mu^+\nu_\mu)= (0.38\pm0.30\pm0.10)\%$.
The measured branching fraction for $D^0\to K^-\mu ^+\nu_\mu$ was previously used
to determine the ratio 
$\Gamma(D^0\to K^-\mu^+\nu_\mu)/ \Gamma(D^+\to \overline K^0\mu^+\nu_\mu)$
combining the previously measured branching fraction 
for $D^+\to \overline K^0\mu^+\nu_\mu$ by the BES Collaboration.
\end{abstract}


\maketitle

\section{\bf Introduction}
The pseudoscalar semileptonic decays $D^0\to K^-\ell^+\nu_\ell$ and $D^0
\to \pi^-\ell^+\nu_\ell$ are the best understood in theory, since the
effects of weak and strong interactions can be separated reasonably well.
Their decay amplitudes are simply related to the Cabibbo-Kobayashi-Maskawa
(CKM) matrix elements $V_{\rm cs}$ and $V_{\rm cd}$, 
which parameterize the mixing between the quark mass
eigenstates and the weak eigenstates, and the form factor describing the
strong interaction between the final state quarks.
As a result, measurements of the branching fractions for $D^0\to K^-\ell^+
\nu_\ell$ and $D^0 \to \pi^-\ell^+\nu_\ell$ play an important role in
understanding of both the weak and strong interactions.
In addition, measurements of the branching fractions for 
$D^0\to K^-\mu^+\nu_\mu$ and  $D^+\to \overline K^0\mu^+\nu_\mu$ can be used to
test isospin symmetry in exclusive semileptonic decays of the charged and
neutral $D$ mesons~\cite{dpk0ev}.

In this paper, we report direct measurements of the branching fractions
for $D^0\to K^-\mu^+\nu_\mu$ and $D^0 \to \pi^-\mu^+\nu_\mu$
(Throughout the paper, charged conjugations
are implied). 
The measured branching fraction for $D^0\to K^-\mu^+\nu_\mu$ 
was previously used to determine the ratio of the partial widths
$\Gamma(D^0\to K^-\mu^+\nu_\mu)/
\Gamma(D^+\to \overline K^0\mu^+\nu_\mu)$~\cite{bes_plbxx_hep_ex_0610020}.
This ratio can be used to test the isospin symmetry in the $D$ semileptonic
decays~\cite{bes_plbxx_hep_ex_0610020}.

\section{\bf BES-II detector}
The BES-II is a conventional cylindrical magnetic detector that is
described in detail in Ref.~\cite{bes}. A 12-layer Vertex Chamber(VC)
surrounding
a beryllium beam pipe provides input to event trigger, as well as
coordinate information. A forty-layer main drift chamber (MDC) located
just outside the VC yields precise measurements of charged
particle trajectories with a solid angle coverage of $85\%$ of 4$\pi$;
it also provides ionization energy loss ($dE/dx$) measurements
for particle identification. Momentum
resolution of $1.7\%\sqrt{1+p^2}$ ($p$ in GeV/$c$) and $dE/dx$
resolution of $8.5\%$ for Bhabha scattering electrons are obtained for
the data taken at $\sqrt{s}=3.773$ GeV. An array of 48 scintillation
counters surrounding the MDC measures time of flight (TOF) of
charged particles with a resolution of about 180 ps for electrons.
Outside the TOF counters, a 12 radiation length, lead-gas barrel shower
counter
(BSC), operating in limited streamer mode, measures the energies of
electrons and photons over $80\%$ of the total solid angle with an
energy resolution of $\sigma_E/E=0.22/\sqrt{E}$ ($E$ in GeV) and spatial
resolutions of $\sigma_{\phi}=7.9$ mrad and $\sigma_Z=2.3$ cm for
electrons. A solenoidal magnet outside the BSC provides a 0.4 T
magnetic field in the central tracking region of the detector. Three
double-layer muon counters instrument the magnet flux return and serve
to identify muons with momentum greater than 500 MeV/c. They cover
$68\%$ of the total solid angle.
 
\section{\bf Data analysis}
The data used in the
analysis were collected with the BES-II detector at the BEPC collider.
A total integrated luminosity of about 33 $\rm pb^{-1}$ was taken at and
around the c.m. (center-of-mass) energy of $\sqrt{s}=3.773$ GeV.
Around the c.m. energy,
the $\psi(3770)$ resonance is produced in electron-positron ($e^+e^-$)
annihilation.
It decays to $D\bar D$ pairs ($D^0\bar D^0$ and $D^+D^-$) 
with a large branching fraction of about $(85 \pm 6)\%$~\cite{bf_psipp_to_dd}.
If a $\bar D^0$ meson is fully reconstructed
(this is called a singly tagged $\bar D^0$ meson) from this data sample,
the $D^0$ meson must exist in the system recoiling against the singly
tagged $\bar D^0$ meson. In the system recoiling against the singly tagged
$\bar D^0$ mesons, we can select the semileptonic decays 
$D^0\to K ^-\mu^+ \nu_\mu$ and $D^0 \to\pi^-\mu^+\nu_\mu$
based on the kinematic signature of the singly tagged $\bar D^0$ event,
and measure branching fractions for these decays directly.

\subsection{\bf \normalsize Event selection}
In order to ensure the well-measured 3-momentum vectors and the
reliability of the charged-particle identification, all charged tracks
are required to be well reconstructed in the MDC with good helix fits, and
to satisfy a geometry cut $|\rm{cos\theta}|<0.85$, where $\theta$ is the
polar angle of the charged track. Each track, except for those from $K^0_S$
decays, must
originate from the interaction region, which is defined by $V_{xy}<2.0$
cm and $|V_z|<20.0$ cm, where $V_{xy}$ and $|V_z|$ are the closest approach
of the charged track in the $xy$-plane and $z$ direction.

Muons, pions and kaons are identified using the $dE/dx$ and TOF measurements,
with which the combined confidence levels ($CL_{\mu}$, $CL_\pi$, or $CL_K$)
for a muon, a pion or a kaon hypotheses are calculated. A pion candidate is
required to have $CL_\pi>0.001$. In order to reduce misidentification,
a kaon candidate is required to satisfy $CL_K>CL_\pi$.
A muon candidate is required to have $CL_{\mu}>0.001$ and
satisfy the relation $CL_{\mu}/(CL_{\mu}+CL_e+CL_K)>0.8$.

Neutral~~kaons~~are~~reconstructed~~through~~the~~decay $K^0_S \to \pi^+
\pi^-$. We require that the $\pi^+$ and the $\pi^-$ 
must originate from a secondary
vertex which is displaced from the event vertex by 4 mm. Moreover,
the difference between the $\pi^+\pi^-$ invariant mass and the $K^0_S$
nominal mass should be less than $20$ MeV/$c^2$.

Neutral~~pions~~are~reconstructed~through the decay $\pi^0 \to \gamma
\gamma$. A good photon candidate must satisfy the following criteria:
(1) the energy deposited in the BSC is greater than 70 MeV;
(2) the electromagnetic shower starts in the first 5 readout layers;
(3) the angle between the photon and the nearest charged track is
greater than $22^\circ$ \cite{dpk0ev,d0kev,epjc1,epjc2};
(4) the opening angle between the direction of the cluster development and
the direction of the photon emission is less than $37^\circ$
\cite{dpk0ev,d0kev,epjc1,epjc2}.

\subsection{\bf \normalsize Singly tagged $\bar D^0$ sample}
The singly tagged $\bar D^0$ sample used in this analysis was selected
in the previous work \cite{d0kev,epjc1}.
The singly tagged $\bar D^0$ mesons are reconstructed in four hadronic
decay modes of $K^+\pi^-$, $K^+\pi^-\pi^-\pi^+$, 
$K^0\pi^+\pi^-$ and $K^+\pi^-\pi^0$. 
The total number of the singly tagged $\bar D^0$ mesons is 
$7584 \pm 198\pm 341$ \cite{d0kev,epjc1}, where
the first error is statistical and the second systematic.

\subsection{\bf \normalsize Candidates for $D^0 \to K^-\mu^+\nu_\mu$ and
$D^0 \to \pi^-\mu^+\nu_\mu$}
Candidates for the semileptonic decays $D^0 \to K^-\mu^+\nu_\mu$ and
$D^0 \to \pi^-\mu^+\nu_\mu$ are selected from surviving tracks in the
system recoiling against the singly tagged $\bar D^0$.
It is required that there are only two oppositely charged tracks, one of
which is identified as muon and the other as kaon or pion.
Muon should have its charge opposite to the charm of the singly tagged
$\bar D^0$ meson, except for the mode $K^0 \pi^+\pi^-$. In addition,
it is required that there should be no isolated photon, which has not been
used in the reconstruction of the singly tagged $\bar D^0$. The
isolated photon should have its energy deposited in the BSC greater than
100 MeV \cite{dpk0ev,d0kev,epjc1,epjc2} and satisfy the photon selection criteria described earlier.

\begin{figure}[hbt]
\includegraphics[width=8.0cm,height=8.0cm]
{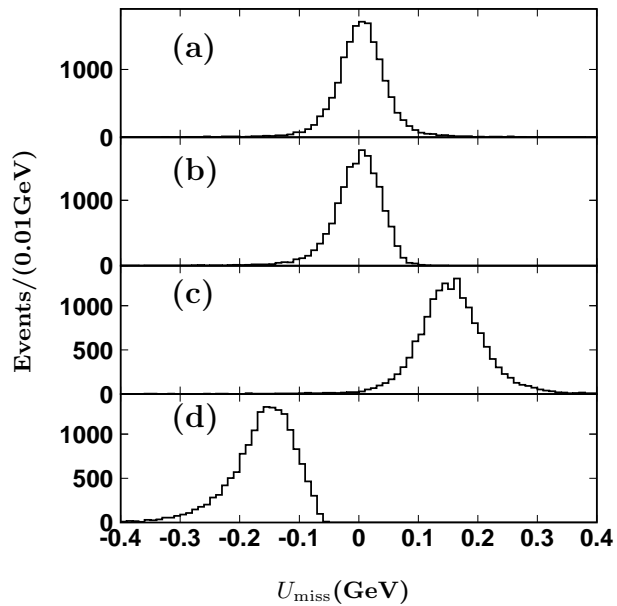}
\put(-140,-5.0){\bf $U_{\rm miss}$(GeV)}
\put(-240,80){\rotatebox{90}{\bf Events/(0.01GeV)}}
\put(-180,200.0){\bf \large (a)}
\put(-180,153.3){\bf \large (b)}
\put(-180,106.6){\bf \large (c)}
\put(-180,59.9){\bf \large (d)}
\caption{
The distributions of the $U_{\rm miss}$ for the Monte Carlo events of
(a) $D^0 \to K^-\mu^+\nu_\mu$;
(b) $D^0 \to \pi^-\mu^+\nu_\mu$;
(c) $D^0 \to K^-\mu^+\nu_\mu$, by replacing kaon mass by pion mass; and
(d) $D^0 \to \pi^-\mu^+\nu_\mu$, by replacing pion mass by kaon mass.
}
\label{umiss_mc}
\end{figure}

In order to obtain the information about the missing neutrino, a kinematic
quantity $U_{\rm miss} \equiv E_{\rm miss} - p_{\rm miss}$ is defined, following the
previous BES works~\cite{dpk0ev,d0kev,epjc1,epjc2}, where $E_{\rm miss}$ and
$p_{\rm miss}$ are the total energy and momentum of all the missing particles.
Figures~\ref{umiss_mc}a and b
show the distributions of the $U_{\rm miss}$ for the Monte Carlo events 
of $D^0 \to K^-\mu^+\nu_\mu$ and 
$D^0 \to \pi^-\mu^+\nu_\mu$.
The values of $\sigma_{U_{\rm miss,i}}$ are $\sim$50 MeV for 
$D^0 \to K^-(\pi^-)\mu^+\nu_\mu$ versus 
$\bar D^0 \to K^+\pi^-\pi^0$, 
and $\sim$40 MeV for $D^0 \to K^-(\pi^-)\mu^+\nu_\mu$ versus the other
singly tagged $\bar D^0$ modes.
To select the candidates for $D^0 \to K^-\mu^+\nu_\mu$ and 
$D^0 \to \pi^- \mu^+\nu_\mu$, 
it is required that each event should have its 
$|U_{\rm miss,i}|<2\sigma_{U_{\rm miss,i}}$. Here, $\sigma_{U_{\rm miss,i}}$ 
is the standard deviation of the $U_{\rm miss,i}$ distribution, obtained by analyzing the
Monte Carlo events of $D^0 \to K^-\mu^+\nu_\mu$ 
(and $D^0 \to \pi^-\mu^+
\nu_\mu$) versus the $i$th singly tagged $\bar D^0$ mode. 
Because kaon can be misidentified as pion, the Cabibbo favored decay
$D^0\to K^-\mu^+\nu_\mu$ can satisfy the selection criteria for the
Cabibbo suppressed decay $D^0\to \pi^-\mu^+\nu_\mu$. In the selection
of $D^0\to \pi^-\mu^+\nu_\mu$, these background events are suppressed
by requiring that $|U_{\rm miss}|<|U_{\pi-as-K}|$, where $U_{\pi-as-K}$ is
calculated by replacing pion mass with kaon mass. 
Figure~\ref{umiss_mc}c shows the distribution of the $U_{\rm miss}$ 
calculated by replacing kaon mass by pion mass
for the Monte Carlo events of $D^0\to K^-\mu^+ \nu_\mu$,
while Fig.~\ref{umiss_mc}d shows the distribution of the
$U_{\rm miss}$ calculated by replacing pion mass by kaon mass 
for the Monte Carlo events of $D^0\to \pi^-\mu^+\nu_\mu$.
The quantity $U_{\rm miss}$ is expected to be closer to zero for the correct
particle assignment.

\begin{figure}[htbp]
\begin{center}
\includegraphics[width=8.0cm,height=8.0cm]
{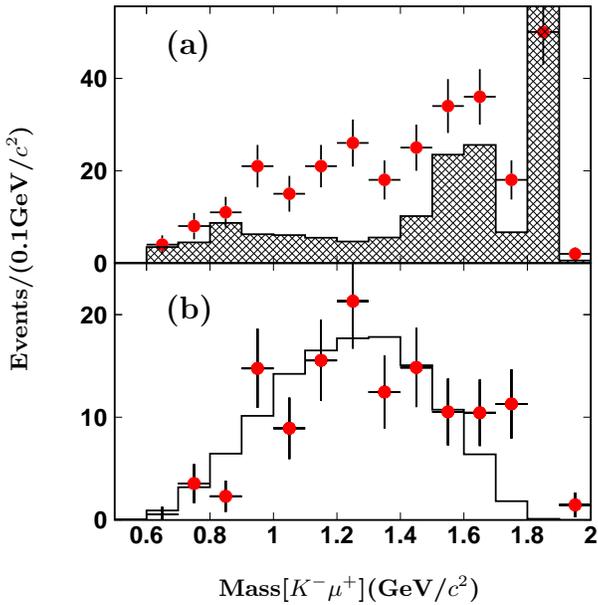}
\put(-160,-5.0){\bf Mass[$K^-\mu^+$](GeV/$c^2$)}
\put(-240,80){\rotatebox{90}{\bf Events/(0.1GeV/$c^2$)}}
\put(-180,200.0){\bf \large (a)}
\put(-180,100.0){\bf \large (b)}
\caption{(a) The distributions of the invariant masses of the $K^-\mu^+$
combinations from the candidates for $D^0 \to K^-\mu^+\nu_\mu$ selected from
the data (point with errors) and from the other decay modes of the $D$
(hatched histogram) selected from Monte Carlo sample;
and (b) the comparison of the background-subtracted mass spectrum of the
$K^-\mu^+$ combinations from the selected candidates
for $D^0 \to K^-\mu^+\nu_\mu$ (point with errors) with the mass spectrum
of the $K^-\mu^+$ combinations from the Monte Carlo events for
$D^0 \to K^-\mu^+\nu_\mu$ (histogram); see text.}
\label{xmkmucut}
\end{center}
\end{figure}

There are possible hadronic backgrounds for the semileptonic decays due
to misidentifying a pion as a muon and due to missing $\pi^0$s.
For example, the hadronic decays $D^0 \to K^-\pi^+$ and 
$D^0 \to K^-\pi^+ \pi^0$ can be misidentified 
as the semileptonic decay $D^0\to K^-\mu^+\nu_\mu$.
Figure \ref{xmkmucut}a shows the distribution of the 
invariant masses of $K^-\mu^+$ combinations
from the candidates for 
$D^0 \to K^- \mu^+\nu_\mu$ 
which satisfy the selection criteria as mentioned above. 
In the figure the point with errors shows the events from the data and the 
hatched histogram represents the expected backgrounds
which are misidentified 
from other $D^0$ decay modes as $D^0 \to K^- \mu^+\nu_\mu$. 
This background shape is estimated by analyzing the Monte Carlo events
for $e^+e^-\to D\bar D$, where the $D$ and $\bar D$ are set to decay into all
possible channels with the branching fractions quoted from PDG~\cite{pdg}.
To examine the mass spectrum of the $K^-\mu^+$ combinations 
from the decay $D^0 \to K^- \mu^+\nu_\mu$, we 
subtract the expected background shape from the mass spectrum
of the $K^-\mu^+$ combinations from the candidates for 
$D^0 \to K^- \mu^+\nu_\mu$ selected from the data.
In Fig.~\ref{xmkmucut}b
the point with errors shows the resulting spectrum of the $K^-\mu^+$
combinations and the histogram shows the mass spectrum of the
$K^-\mu^+$ combinations from the Monte Carlo events 
of $D^0 \to K^- \mu^+\nu_\mu$.
The mass spectra of $K^-\mu^+$ combinations 
from both the data and the Monte Carlo events matching well indicates
that we can correctly count the number of the signal events for the decay
$D^0 \to K^-\mu^+\nu_\mu$ from the data 
after subtracting the number of the background events 
from other decay modes of the $D^0$ mesons.   
Monte Carlo study shows that the dominant
background events are from the hadronic decays $D^0 \to K^-\pi^+$
and $D^0 \to K^-\pi^+\pi^0$.
To effectively remove these background events, 
we require that the invariant masses of the $K^-\mu^+$ combinations
from the selected candidates to be less than 1.60 GeV/$c^2$.
Similarly, in the selection of $D^0 \to \pi^-\mu^+\nu_\mu$, 
the invariant masses of the $\pi^-\mu^+$ combinations
are also required to be less than 1.60 GeV/$c^2$.

\begin{figure}[hbtp]
\begin{center}
\includegraphics[width=8.0cm,height=8.0cm]
{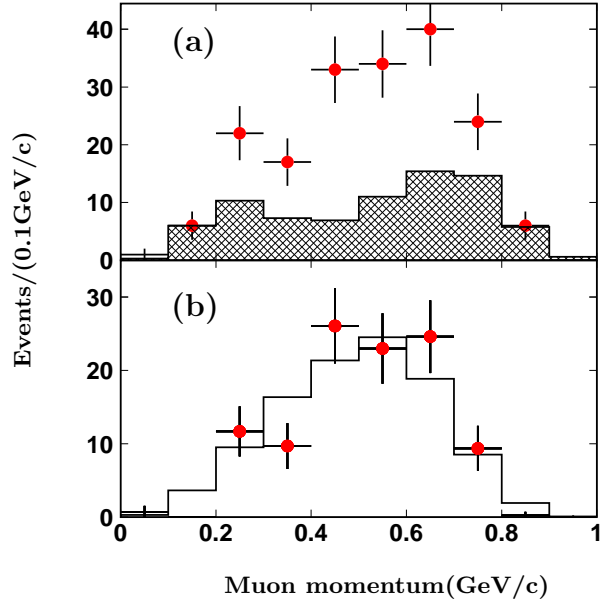}
\put(-160,-5.0){\bf Muon momentum(GeV/c)}
\put(-240,80){\rotatebox{90}{\bf Events/(0.1GeV/c)}}
\put(-180,200.0){\bf \large (a)}
\put(-180,100.0){\bf \large (b)}
\caption{(a) The distributions of the muon momenta of the
candidates for $D^0 \to K^-\mu^+\nu_\mu$ selected from
the data (point with errors) and from
other decay modes of $D^0$ from Monte Carlo sample (hatched histogram);
and (b) comparison of the background-subtracted momentum spectrum
(point with errors)
and the spectrum from the Monte Carlo events of
$D^0 \to K^-\mu^+\nu_\mu$ (histogram).}
\label{xpmuon}
\end{center}
\end{figure}

Figure~\ref{xpmuon}a shows the distributions of the momentum of the muon
from the candidates for $D^0 \to K^-\mu^+\nu_\mu$ 
selected from the data (point with errors) and from other decay modes (hatched
histogram) of the $D^0$ which are obtained from the Monte Carlo samples 
for $e^+e^- \to D^0 \bar D^0$.
Subtracting the estimated background shape of the Monte Carlo events
from the muon momentum spectrum yields the expected momentum distribution of the
muon from the semileptonic decays $D^0 \to K^-\mu^+\nu_\mu$.
In Fig.~\ref{xpmuon}b the point with errors shows the background-subtracted
muon momenta and the histogram shows the distribution of the muon momenta
from the Monte Carlo events for $D^0 \to K^-\mu^+\nu_\mu$.
The expected momentum distribution of the muons from the data agrees well with
the one from the Monte Carlo events for $D^0 \to K^-\mu^+\nu_\mu$.

\begin{figure}
\begin{center}
 \includegraphics[width=8.0cm,height=8.0cm]
{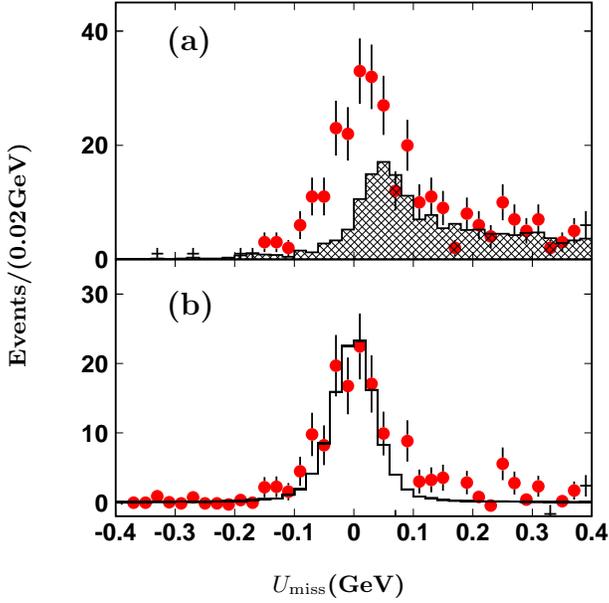}
\put(-140,-5.0){\bf $U_{\rm miss}$(GeV)}
\put(-240,80){\rotatebox{90}{\bf Events/(0.02GeV)}}
\put(-180,200.0){\bf \large (a)}
\put(-180,100.0){\bf \large (b)}
\caption{
The distributions of the $U_{\rm miss}$ of the candidates for $D^0 \to
K^-\mu^+\nu_\mu$,
(a) the point with errors show the events selected from the data,
while the hatched histogram represents the expected background
from $D\bar D$ Monte Carlo events; and (b) comparison of the data
(point with errors) after subtracting the background and the Monte Carlo
events of
$D^0 \to K^-\mu^+\nu_\mu$ (histogram).}
\label{umiss_compare}
\end{center}
\end{figure}

To further check the selected candidates for the decay 
$D^0 \to K^-\mu^+\nu_\mu$, we examine the distributions of the quantity
$U_{\rm miss}$. We open the criterion $|U_{\rm miss,i}|<2\sigma_{U_{\rm miss,i}}$
in the criteria for selection of the events for the decay $D^0 \to K^-\mu^+\nu_\mu$. 
Figure~\ref{umiss_compare}a shows the distribution of the $U_{\rm miss}$
calculated for the
candidates for $D^0 \to K^-\mu^+\nu_\mu$ 
versus the singly tagged $\bar D^0$ mesons selected from the data
(point with errors), and from other decay modes of the Monte Carlo
events (the hatched histogram) for $e^+e^- \to D\bar D$.
Subtracting the background (hatched histogram) from the distribution of the
$U_{\rm miss}$ of the candidates for $D^0 \to K^-\mu^+\nu_\mu$ selected from the
data yields the expected distribution of the $U_{\rm miss}$
for the decay $D^0 \to K^-\mu^+\nu_\mu$ (point with errors)
as shown in Fig.~\ref{umiss_compare}b.
In Fig.~\ref{umiss_compare}b, the histogram shows
the distribution of the $U_{\rm miss}$ calculated for the Monte Carlo events of
$D^0 \to K^-\mu^+\nu_\mu$ versus 
the singly tagged $\bar D^0$.

\begin{figure}
\begin{center}
 \includegraphics[width=8.0cm,height=8.0cm]
{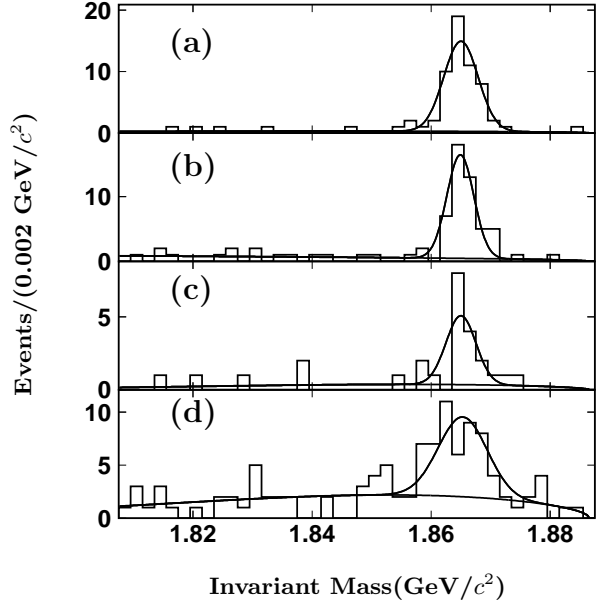}
\put(-165,-5){\bf Invariant Mass(GeV/$c^2$)}
\put(-240,70){\rotatebox{90}{\bf Events/(0.002 GeV/$c^2$)}}
\put(-180,200.0){\bf \large (a)}
\put(-180,153.3){\bf \large (b)}
\put(-180,106.6){\bf \large (c)}
\put(-180,59.9){\bf \large (d)}
\caption{
The distributions of the fitted $Kn\pi$ invariant masses for the events,
selected from the data, in which the candidates for $D^0 \to K^-\mu^+
\nu_\mu$ are observed in the system recoiling against the singly tagged
$\bar D^0$
in decay modes of (a) $K^+\pi^-$, (b) $K^+\pi^-\pi^-\pi^+$,
(c) $K^0\pi^+ \pi^-$ and (d) $K^+\pi^-\pi^0$.}
\label{d0_kmunu}
\end{center}
\end{figure}

\begin{figure}
\begin{center}
 \includegraphics[width=8.0cm,height=4.0cm]
{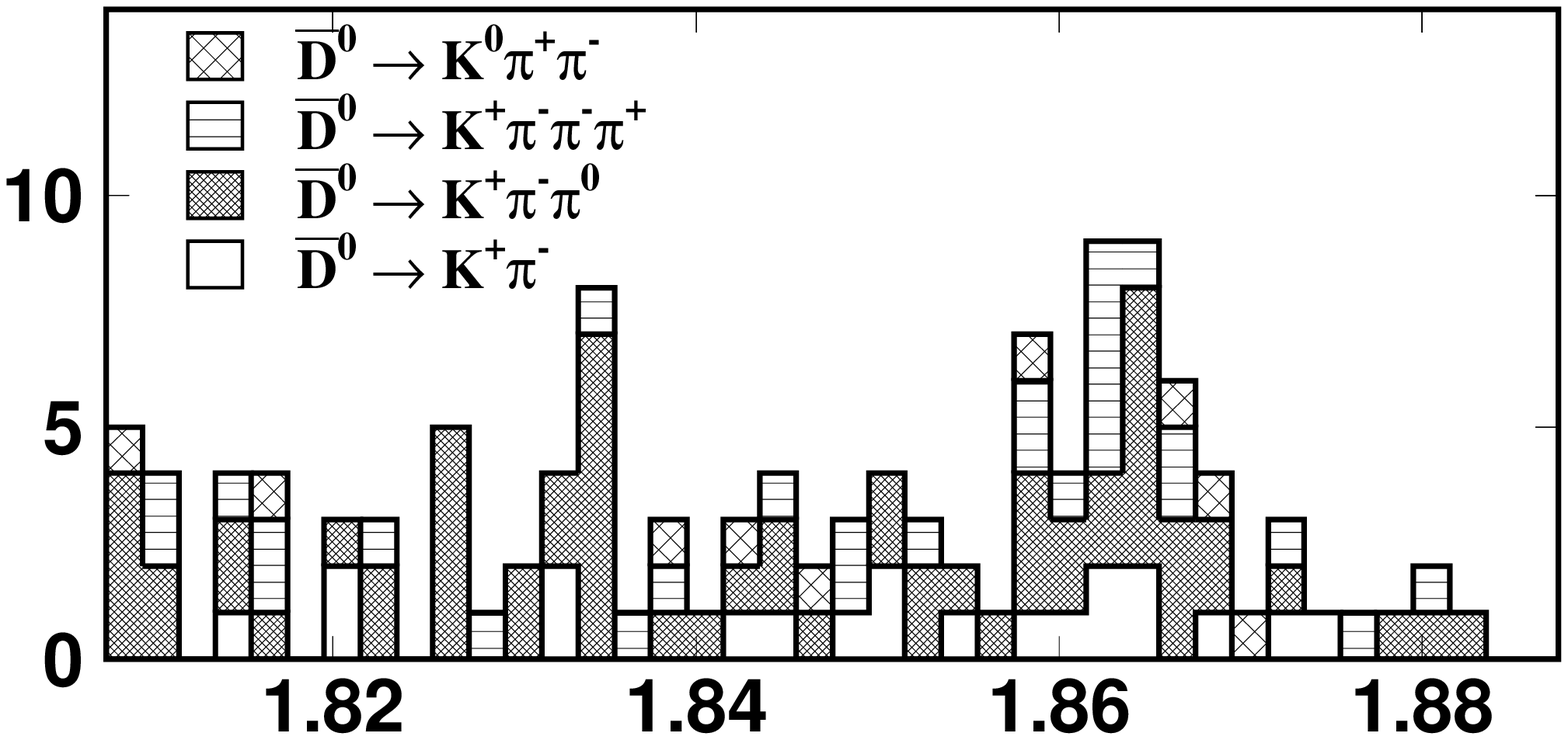}
\put(-165,-5){\bf Invariant Mass(GeV/$c^2$)}
\put(-240,5){\rotatebox{90}{\bf Events/(0.002 GeV/$c^2$)}}
\caption{
The distribution of the fitted $Kn\pi$ invariant masses for the events,
selected from the data, in which the candidates for $D^0 \to
\pi^-\mu^+\nu_\mu$
are observed in the system recoiling against the singly tagged $\bar D^0$.}
\label{d0_pimunu}
\end{center}
\end{figure}

Figures~\ref{d0_kmunu} and \ref{d0_pimunu}
show, respectively, the distributions of the invariant masses of the
$Kn\pi$ (i.e. $K^+\pi^-$ or $K^+\pi^-\pi^-\pi^+$ or
$K^0\pi^+\pi^-$ or $K^+\pi^-\pi^0$)
combinations
from the events in which the candidates for $D^0\to K^-\mu^+ \nu_\mu$ 
and $D^0\to\pi^-\mu^+\nu_\mu$ are observed in the system recoiling against 
the $\bar D^0$ tags. Fitting to the mass spectra as shown in Fig.~\ref{d0_kmunu}
with a Gaussian function for the $\bar D^0$ signal and a special background
function~\cite{dpk0ev,d0kev} to describe background
yields $152.5\pm13.6$ candidates for $D^0 \to K^-\mu^+\nu_\mu$. 
In Fig.~\ref{d0_pimunu}, there are 45 events in the $\pm 3\sigma_{M_{D_i}}$ mass
window around the fitted $\bar D^0$ meson mass $M_{D_i}$, 
and 74 events in the outside of the signal regions, 
where $\sigma_{M_{D_i}}$ is the standard deviation of the 
distribution of the $Kn\pi$ combinations 
for the $i$th mode. By assuming that the background
distribution is flat except the one described in subsection \ref{back}, 
$25.8\pm3.1$ background events are estimated in the
signal regions. Subtracting the number of background events, 
we obtain $19.2 \pm7.4$ candidates for $D^0 \to \pi^-\mu^+\nu_\mu$.

\subsection{\bf \normalsize Background subtraction}
\label{back}
The events from other hadronic or semileptonic decays may also satisfy the
selection criteria for the semileptonic decays and are misidentified as the
$D^0\to K^-\mu^+\nu_\mu$ and $D^0\to \pi^-\mu^+\nu_\mu$ decay events. The
peak-like background due to these misidentified events can not be
illustrated by the background shape as mentioned earlier. The numbers of
these background events have to be subtracted from the candidates for
the semileptonic decays. 
The number of the backgrounds
are estimated by analyzing a Monte Carlo sample which is
about fourteen times larger than the data. The Monte Carlo events are
generated as $e^+e^- \to D\bar D$, where the $D$ and $\bar D$ mesons are
set to decay into all possible final states with the branching fractions
quoted from PDG~\cite{pdg} excluding the two decay modes under study. The
numbers of the events satisfying the selection criteria are then normalized
to the corresponding data set.
Monte Carlo study shows that the dominant background for 
$D^0 \to K^-\mu^+\nu_\mu$ is from the hadronic decay $D^0 \to K^-\pi^+\pi^0$,
and the main backgrounds for $D^0 \to \pi^-\mu^+\nu_\mu$ are from
$D^0\to \pi^+\pi^-\pi^0$, $D^0\to K^-\pi^+\pi^0$, $D^0\to K^-\pi^+$,
$D^0\to K^0\pi^+\pi^-$ and $D^0 \to K^-\mu^+\nu_\mu$.
Totally $65.3\pm 2.5$ and $9.9\pm 1.2$ background events are obtained
for $D^0\to K^-\mu^+\nu_\mu$ and $D^0 \to \pi^-\mu^+ \nu_\mu$, respectively.
After subtracting these numbers of background events, 
$87.2 \pm 13.6$ and $9.3 \pm 7.4$ 
signal events for $D^0\to K^-\mu^+\nu_\mu$ 
and $D^0\to\pi^-\mu^+\nu_\mu$ decays are retained, respectively.

\section {\bf Results}

The measured branching fractions for the semileptonic decays
are obtained by dividing the observed numbers of the semileptonic decay
events $N[D^0 \rightarrow K^-(\pi^-) \mu^+ \nu_{\mu}]$ 
by the number $N_{\bar D^0_{\rm tag}}$ of the
singly tagged $\bar D^0$ mesons and the
reconstruction efficiencies $\epsilon_{K^-(\pi^-) \mu^+ \nu_{\mu}}$,
\begin{small}
\begin{equation}
BF(D^0 \rightarrow K^-(\pi^-) \mu^+ \nu_{\mu}) = \\
\frac{N[D^0 \rightarrow K^-(\pi^-) \mu^+ \nu_{\mu}]}
{\epsilon_{K^-(\pi^-) \mu^+ \nu_{\mu}}
  N_{\bar D^0_{\rm tag}}}.
\label{brsemi}
\end{equation}
\end{small}
The efficiencies for reconstruction of the semileptonic decay events
of $D^0 \rightarrow K^- \mu^+ \nu_{\mu}$ and 
$D^0 \rightarrow \pi^- \mu^+ \nu_{\mu}$
are estimated by Monte Carlo simulation.
A detail Monte Carlo study gives that the efficiencies are
$\epsilon_{K^- \mu^+ \nu_{\mu}}=(32.39\pm 0.24)\%$ and
$\epsilon_{\pi^- \mu^+ \nu_{\mu}}=(32.04\pm 0.25)\%$.
Inserting these numbers in (\ref{brsemi}), we obtain the branching
fractions for the semileptonic decays
$$BF(D^0\to K^-\mu^+\nu _\mu) = (3.55\pm0.56\pm0.59)\% $$
and 
$$BF(D^0 \to \pi^- \mu^+\nu_\mu)= (0.38\pm0.30\pm0.10)\%, $$
where the first error is statistical and the second systematic.
In the measurements of the branching fractions, the systematic error
arises mainly from the uncertainties in tracking efficiency ($\sim$2.0\%
per track), in particle identification ($\sim$0.5\% per charged kaon or
pion, $\sim$1.5\% per muon), in photon selection ($\sim$2.0\%), in
$U_{\rm miss}$ selection ($\sim$0.6\% for $D^0\to K^-\mu^+\nu_\mu$ and $\sim$
1.2\% for $\pi^- \mu^+\nu_\mu$), in the number of the singly tagged $\bar
D^0$ mesons ($\sim$4.5\%), in Monte Carlo statistics ($\sim$0.7\% for
$D^0\to K^-\mu^+\nu_\mu$ and $\sim$0.8\% for $D^0 \to \pi^-\mu^+\nu_\mu$),
in background fluctuation ($\sim$2.9\% for $D^0\to K^-\mu^+\nu_\mu$,
$\sim$12.9\% for $D^0 \to \pi^- \mu^+\nu_\mu$) and 
in background estimation ($\sim$15.0\% for $D^0\to K^-\mu^+\nu_\mu$, 
$\sim$21.3\% for $D^0 \to \pi^- \mu^+\nu_\mu$) due to
unknown branching fractions of some background channels.
Adding these uncertainties in quadrature yields the total systematic
errors of $\sim$16.7\% for $D^0\to K^-\mu^+\nu_\mu$ and $\sim$25.8\% for
$D^0 \to \pi^-\mu^+\nu_\mu$.

Table~\ref{compare} gives a comparison of the measured branching fractions
for $D^0\to K^-\mu^+ \nu_\mu$ and $D^0\to \pi^-\mu^+ \nu_\mu$ 
by the BES Collaboration in this work, along
with those measured by other Collaborations~\cite{cleo,e687,e687_2,e653,focus}.
For the measurements from other experiments, 
the branching fraction for $D^0\to K^-\mu^+\nu_\mu$
(or $D^0\to \pi^-\mu^+\nu_\mu$) was measured 
relative to the decay $D^0 \to K^-\pi^+$ or $D^0\to\mu^+X$ 
(or $D^0\to K^-\mu^+\nu_\mu$). The branching fraction 
for $D^0\to K^-\mu^+\nu_\mu$ (or $D^0\to \pi^-\mu^+\nu_\mu$) is
calculated by multiplying the measured ratio
$\Gamma(D^0\to K^-\mu^+ \nu_\mu)/\Gamma(D^0\to K^-\pi^+)$ or 
$\Gamma(D^0\to K^-\mu^+\nu_\mu)/\Gamma(D^0\to \mu^+X)$ 
(or $\Gamma(D^0\to \pi^-\mu^+\nu_\mu)/\Gamma(D^0\to K^-\mu^+\nu_\mu)$) 
by the branching fraction for $D^0\to K^-\pi^+$ or
$D^0\to \mu^+X$ (or $D^0\to K^-\mu^+ \nu_\mu$) from PDG \cite{pdg}. 
The directly measured branching fractions reported in this paper
agree within error with those measured by the CLEO, E687, E653 and FOCUS
Collaborations~\cite{cleo,e687,e687_2,e653,focus}.

\begin{table*}[htbp]
\caption{\normalsize Summary of the measured branching fractions for
$D^0\to K^-\mu^+\nu_\mu$ and $D^0\to \pi^-\mu^+\nu_\mu$ from different
experiments.}
\begin{center}
\begin{tabular}{c|l|l} \hline
Collab.          & $BF(D^0\to K^-\mu^+\nu_\mu)(\%)$
                 & $BF(D^0\to \pi^-\mu^+\nu_\mu)(\%)$ \\ \hline
BES              & $3.55\pm0.56\pm0.59$       &$0.38\pm0.30\pm0.10$\\
CLEO\cite{cleo}  & $3.00\pm0.30\pm0.34\pm0.06$& \\
E687\cite{e687}  & $3.12\pm0.49\pm0.49\pm0.06$& \\
E687\cite{e687_2}& $3.24\pm0.13\pm0.11\pm0.06$& \\
E653\cite{e653}  & $3.07\pm0.33\pm0.26\pm0.33$& \\
FOCUS\cite{focus}& & $0.24\pm0.03\pm0.02\pm0.01$ \\ \hline
\end{tabular}
\label{compare}
\end{center}
\end{table*}

With~~the~~measured~~branching~~fractions~~for~~$D^0 \to K^-\mu^+\nu_\mu$ and
$D^0 \to \pi^-\mu^+\nu_\mu$, we determine the ratio of the two branching fractions
\begin{equation}
\frac{BF(D^0 \to \pi^-\mu^+\nu_\mu)}
{BF(D^0 \to K^-\mu^+\nu_\mu)}= 0.11\pm0.09\pm0.03,
\end{equation}
where the first error is statistical and the second systematic,
which arises mainly from the uncanceled uncertainties including
$U_{\rm miss}$ selection ($\sim$1.3\%), Monte Carlo statistics ($\sim$1.1\%),
and background subtraction ($\sim$29.1\%). 

Our measured branching fraction for $D^0 \to K^-\mu^+\nu_\mu$
was previously used to determine the ratio of the partial widths
$\Gamma(D^0\to K^-\mu^+\nu_\mu)/ \Gamma(D^+\to \overline K^0\mu^+\nu_\mu)
=0.87\pm0.24\pm0.15$~\cite{bes_plbxx_hep_ex_0610020}.
Combining the ratio with the one obtained by analyzing the semielectronic
modes yields the ratio $\Gamma(D^0\to K^-\ell^+\nu_\ell)/ \Gamma(D^+\to
\overline
K^0\ell^+\nu_\ell) = 1.00\pm 0.17 \pm 0.06$~\cite{bes_plbxx_hep_ex_0610020},
which is in good agreement with $\Gamma(D^0\to K^-\ell^+\nu_\ell)/ \Gamma(D^+\to
\overline K^0\ell^+\nu_\ell) = 1$ predicted by the
isospin symmetry in the exclusive semileptonic decays of the charged and
neutral $D$ mesons.

\section{\bf Summary}
In summary, using the data sample of about 33 pb$^{-1}$ collected at and
around 3.773 GeV with the BES-II detector at the BEPC collider, the absolute
branching fractions for the decays $D^0\to K^-\mu^+\nu_\mu$ and $D^0\to
\pi^-\mu^+\nu_\mu$ have been measured. These are
$BF(D^0\to K^-\mu ^+\nu_\mu)= (3.55\pm0.56\pm0.59)\%$ 
and $BF(D^0\to\pi^-\mu^+\nu_\mu)= (0.38\pm0.30\pm0.10)\%$. 
The ratio of the branching fractions for the two decays is determined to be 
$BF(D^0 \to \pi^-\mu^+\nu_\mu)/BF(D^0 \to K^-\mu^+\nu_\mu)=0.11\pm 0.09\pm 0.03$.
The measured branching fraction for $D^0\to K^-\mu ^+\nu_\mu$ was previously used
to determine the ratio
$\Gamma(D^0\to K^-\mu^+\nu_\mu)/ \Gamma(D^+\to \overline K^0\mu^+\nu_\mu)$
combining the previously measured branching fraction for 
$D^+\to \overline K^0\mu^+\nu_\mu$~\cite{bes_plbxx_hep_ex_0610020}
by the BES Collaboration. 

\section {\bf Acknowledgements}
The BES collaboration thanks the staff of BEPC and computing center
for their hard
efforts. This work is supported in part by the National Natural
Science Foundation of China under contracts Nos. 10491300,
10225524, 10225525, 10425523, the Chinese Academy of Sciences under
contract No. KJ 95T-03, the 100 Talents Program of CAS under
Contract Nos. U-11, U-24, U-25, and the Knowledge Innovation
Project of CAS under Contract Nos. U-602, U-34 (IHEP), and the
National Natural Science Foundation of China under Contract No.
10225522 (Tsinghua University).

\end{document}